\documentclass[a4paper, 11pt] {report}
\usepackage{amsmath}
\usepackage{fullpage}
\usepackage[latin1]{inputenc}
\usepackage[T1]{fontenc}
\usepackage{graphicx}
\usepackage{enumerate}
\usepackage{nomencl}
\usepackage{import}
\usepackage{amsmath}
\usepackage{subcaption}
\usepackage{color}

\newcommand{\com}[2]{\left\lbrack{#1,#2}\right\rbrack}
\newcommand{\ket}[1]{\left| #1 \right>}

\newcommand{\braket}[2]{\left<{ #1 \left| #2 \right.}\right>}

\begin{document}
\title{Nonlinear strong coupling between sub-band excitons:\\ a new coherent path for ultrafast energy relaxation}

\author{Pierre Gilliot, Bertrand Yuma, Marc Ziegler, Mathieu Gallart, Bernd H{\"o}nerlage}
\maketitle

\section*{Abstract}
We demonstrate theoretically and experimentally that the nonlinear interaction between excitations whose harmonic energies coincide gives rise to a strong coupling that opens a new coherent ultrafast energy relaxation path. Instead of an incoherent decay of excitations, that should take a finite time that depends on the energy difference between the initial state and the final state, the nonlinear interaction allows their coherent superposition and thus an instantaneous transfer of the excitation over energies as large as the electron-volt. Such a situation should be encountered in many systems. We demonstrate that such a model applies also for excitons in single-wall carbon nanotubes (SWCNT) where a strong nonlinear Coulomb interaction occurs between $\rm E_{11}$ and $\rm E_{22}$ states. This explains a wide panel of observations about  optoelectronic properties of the SWCNT and gives a coherent picture of their features like the exciton-line spectral positions, exciton collisions and their ultrafast relaxation, as well as the low light-emission efficiency of the nanotubes.

\section*{Introduction}
Very rapidly when semiconducting single-wall carbon nanotubes (SWCNT) started to be investigated, their optoelectronic properties were shown to be be strongly dominated by the effects of Coulomb interactions between photoexcited carriers. The sharp lines that were observed in emission and absorption, first assigned to van Hove singularities that are 1D features, were further explained as excitonic transitions~\cite{kane_ratio_2003}. The strength of the Coulomb interaction is so important that the binding energy of those electron-hole pairs is as large as sizeable part of the gap energy, in the range of an electronvolt fraction. Other multicarrier complexes have been more recently identified, like biexcitons~\cite{Yuma2013, colombier_2012} and trions \cite{matsunaga_observation_2011, Santos2011}. Both show very strong binding energies, around a few hundreds of meV. 

\begin{figure}[h!]
	\begin{center}
		\includegraphics[scale=0.5]{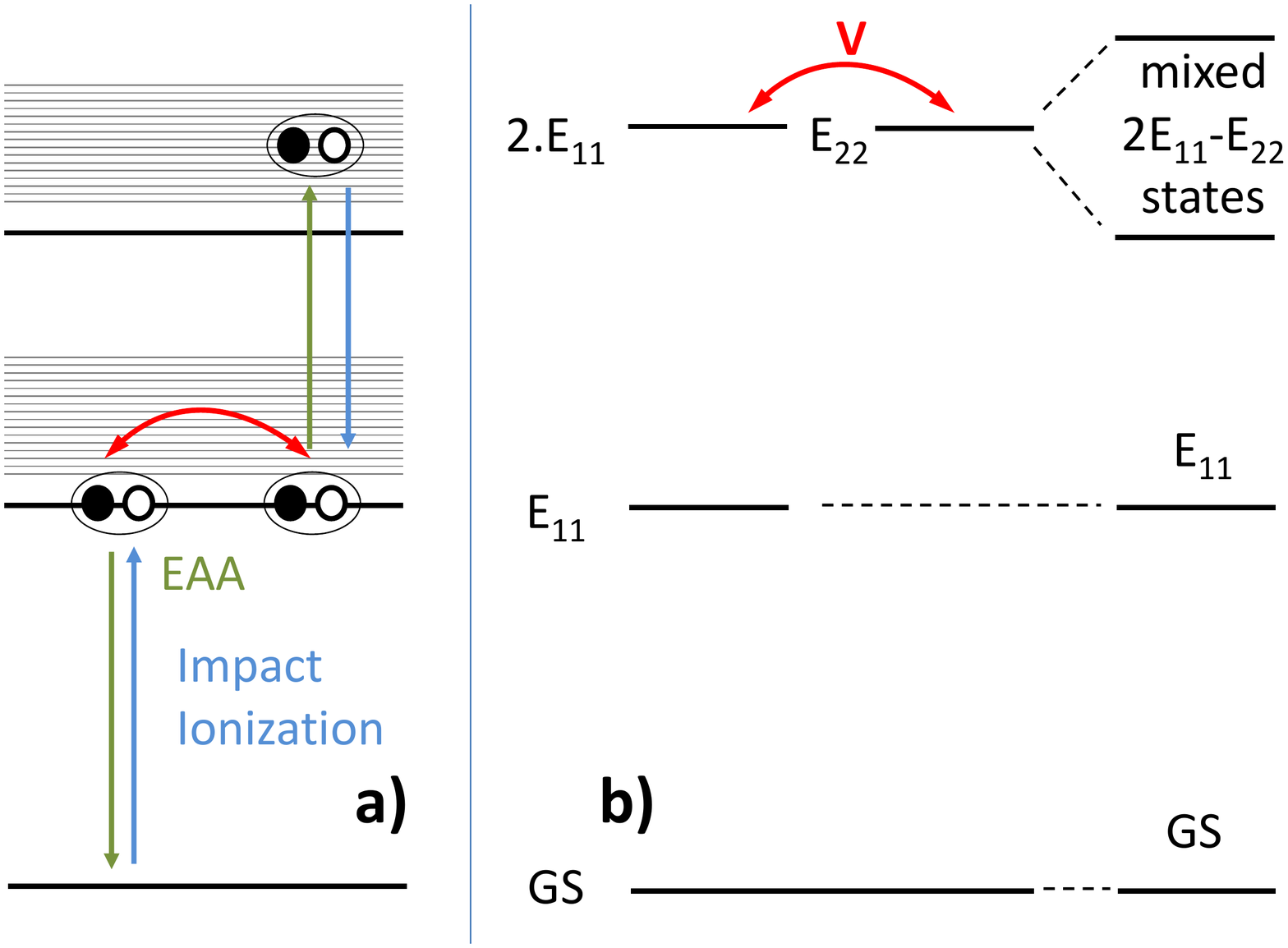}
		\caption {a) Scheme of the usual EEA mechanism: the collision between two excitons gives rise to the annihilation of one of them and to the transfer of the other to a higher energy level. In the reverse process, that is similar to Impact Ionization, one exciton dissociates into two excitons. b) Proposed mechanism where strong coupling between 2$\rm E_{11}$ and $\rm E_{22}$ two-exciton states leads to the formation of mixed states.}
		\label{Fig_strongcoupling}
	\end{center}
\end{figure}

On the other side, measurements of the exciton dynamics did show, from the very first experiments which were performed~\cite{wang_observation_2004, valkunas_exciton-exciton_2006}, that the main population-decay mechanism is also due to Coulomb interaction~\cite{Langlois2015}: exciton-exciton annihilation (EEA) is the collision~\cite{NguyenPRL2011} between two excitons that gives rise to the annihilation of one of them, whose energy is transfered to the other one (Fig.~\ref{Fig_strongcoupling}a). The measured decays were thus typical for a non-radiative recombination. However, EEA is an Auger-like process, that is also based on Coulomb interaction between particles. Auger relaxation is linked to a reverse mechanism called "impact ionization", where one particle dissociates and shares its energy into two particles. But Auger process and impact ionization, as well as EEA and its counterpart, become energy relaxation processes only because they are followed by a rapid transfer of the particles towards available continua of states. If this is not the case, the initial and final states of the interactions remain coherent: that would give rise to their strong coupling and to the formation of mixed states (Fig.~\ref{Fig_strongcoupling}b). 

We will show in this publication that this situation occurs for $\rm E_{11}$ and $\rm E_{22}$ subband excitons in SWCNT. Indeed, because of the linear dispersion of electronic states in graphen, the first two subbands in nanotubes have nearly an energy ratio of two, if the so-called trigonal wrapping is neglected. The energy of the state of two  $\rm E_{11}$ excitons is thus nearly resonant with the energy of a $\rm E_{22}$ excitons. These states can then be strongly coupled by the Coulomb interaction between excitons. Subband exciton states in carbon nanotubes are thus a model of choice to demonstrate the nonlinear coupling of excitations that gives rise to the formation of strongly coupled state and the opening of a very efficient and ultrafast relaxation path. 

We will here present experimental results about exciton dynamics and show that they can be explained by a strong coupling between $\rm E_{11}$ and $\rm E_{22}$ subband excitons. In a first part, we present a simple model that extends the EEA relaxation process towards a reversible strong coupling mechanism between $\rm E_{11}$ and $\rm E_{22}$ resonances. Then we demonstrate that it gives rise to an ultrafast energy relaxation path for the excitons. In a second part, we present our experiments and results: we show that they demonstrate collision processes between excitons with a finite size and we clarify incidentally some usual misunderstandings about the different typical lengths that characterize exciton.  In a last part, we show that many features evidenced in the optical studies of SWNT can be clarified and gathered in an unified picture of their properties. 

\section*{Strong coupling of excitons}
\begin{figure}[h!]
	\begin{center}
		\includegraphics[scale=0.5]{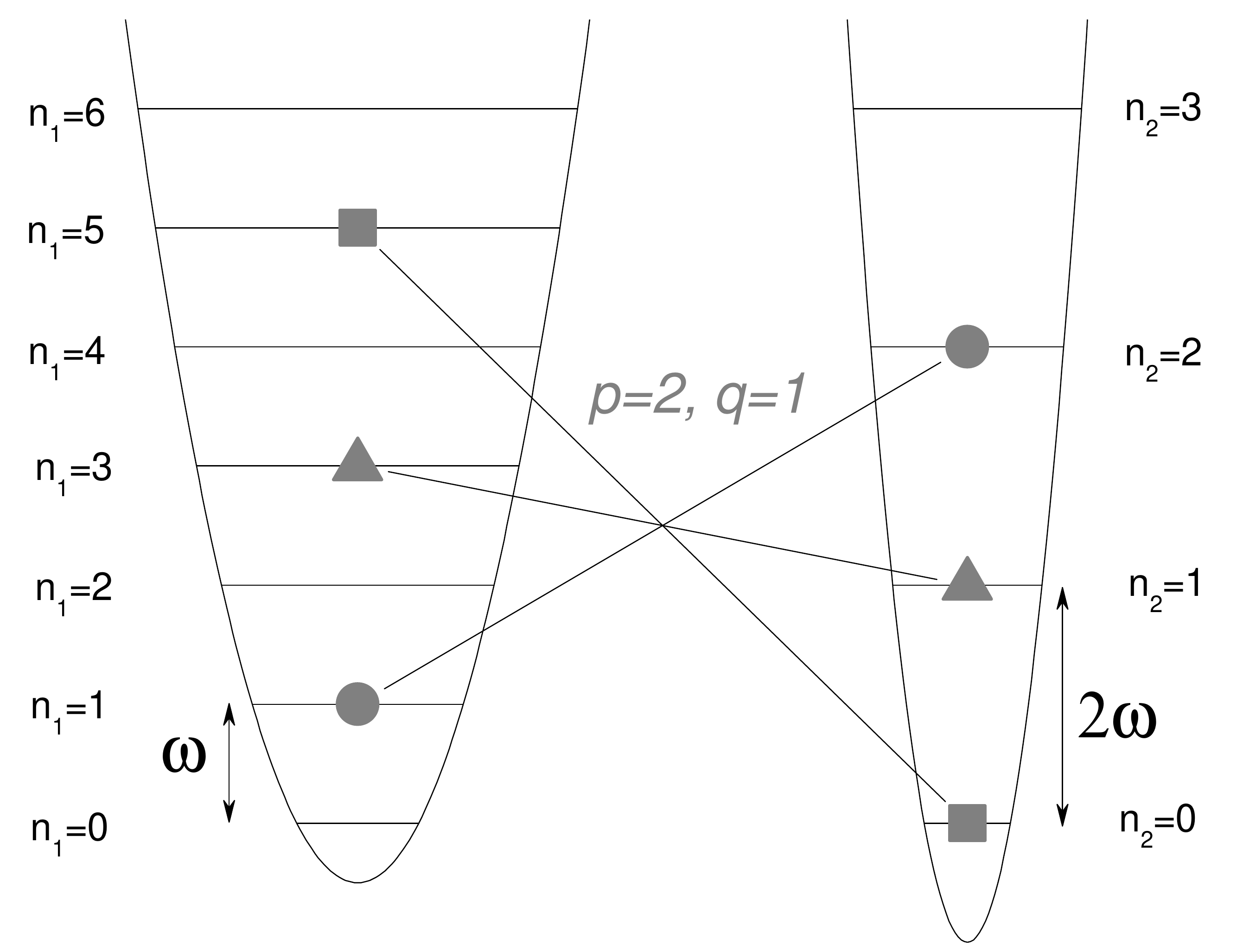}
		\caption {Energy levels of two harmonic oscillators with energies $E_1=\hbar \omega$ and $E_2=2\hbar \omega$. An example is given of the effect of the nonlinear interaction that couples states that have nearly the energy $5\omega$: $\left|n_1=1, n_2=2>\right\rangle$ (circles), $\left|n_1=3, n_2=1>\right\rangle$ (triangles)\textbf{}, and $\left|n_1=5, n_2=0>\right\rangle$ (squares).}
		\label{Fig_oscLevels}
	\end{center}
\end{figure}

\begin{figure}[h!]
	\begin{center}
		\includegraphics[scale=0.25]{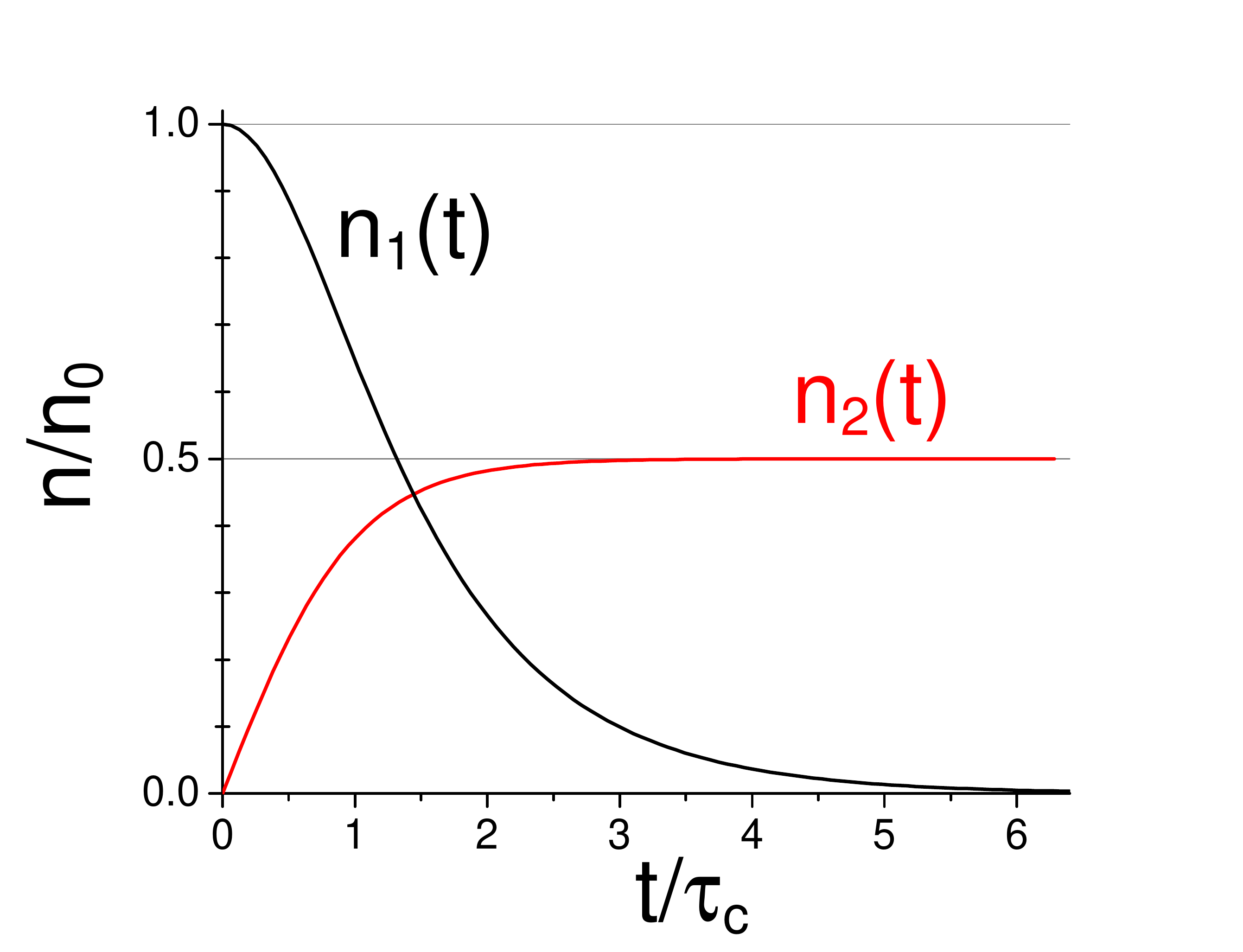}
		\caption {Populations, as a function of the time of two harmonic oscillators, with energies $E_1=\hbar \omega$ and $E_2=2\hbar \omega$ that are nonlinearly coupled in the classical limit where the number of quanta is very high. For long times, there is a full population transfer towards the $E_2$ oscillator.}
		\label{Fig_OscClassiques}
	\end{center}
\end{figure}

In a first approximation, excitons can be considered as quasi bosons~\cite{HaugSchmittRink1984} (see below). In a simplified picture, the interaction between $\rm E_{11}$ and $\rm E_{22}$ excitons, with energies $E_1$ and $E_2\simeq 2.E_1$, respectively, can thus be reduced to the case of two harmonic oscillators with energy ratio of 2 (Fig.~\ref{Fig_oscLevels}), that interact nonlinearly. 
The Hamiltonian is then given by:
\begin{eqnarray}
{\hat H}^{(NL)}
= 
E_1\, \left({\hat a}_1^\dag {\hat a}_1 +{1\over 2}\right)
+E_2\, \left({\hat a}_2^\dag {\hat a}_2 +{1\over 2}\right)
+G^*\,{\hat a}_1^\dag{\hat a}_1^\dag{\hat a}_2
+G\,{\hat a}_1{\hat a}_1{\hat a}_2^\dag	
\label{NLhamiltonian}
\end{eqnarray}
${\hat a}_j^\dag$ and ${\hat a}_j$ being creation and annihilation operators. 
In the interaction term, two $E_1$ quanta are annihilated (resp. created) when one $E_2$ quantum is created (resp. annihilated) (Fig.~\ref{Fig_oscLevels}). 
For the operators in the Heisenberg representation, written $\check{a}_j (t)$, the equations of evolution are given by
\begin{align}
	i\hbar {d\check{a}_1 (t)\over dt} 
	& =
	\com{\check{a}_1(t)}{\check{H}^{(NL)}(t)}
	=
	E_1 \check{a}_1(t)
	-2 G^*\,{\check a}_1^\dag(t) {\check a}_2(t)
	\nonumber
\\
	i\hbar {d\check{a}_2 (t)\over dt} 
	& =
	\com{\check{a}_2(t)}{\check{H}^{(NL)}(t)}
	=
	E_2 \check{a}_2(t)
	+ G\,{\check a}_1^2(t) 
	\label{dynNL}
\end{align}
At large number of excitons, the operators $\check{a}_j (t)$ become C-numbers  that are the normal variables $\alpha_j (t)$ of the oscillating modes. They are related to the number of quanta $n_j=\left|\alpha_j\right|^2 $. Those equations can be solved using perturbation theory. Starting from the modes of the uncoupled oscillators, we write
\begin{align}
\alpha_j (t) &= \frac{A_j(t)}{\sqrt{E_j/\hbar}} e^{-{\rm i}E_j t/\hbar}
\end{align}
Keeping only terms that oscillate at the same frequency and neglecting second order derivatives, one obtains: 
\begin{align}
	{d A_1(t) \over dt}
	&=
	-{\rm i}\kappa^*\,
	A_2(t) A_1^*(t) e^{{\rm i}\left(E_2-2E_1\right) t/\hbar}
	\\
	{d A_2(t) \over dt}
	&=
	- {\rm i}\kappa\,{E_2\over 2E_1}
	A_1^2(t)e^{-{\rm i}\left(E_2-2E_1\right) t/\hbar}
\end{align}
\begin{equation}
\kappa
={12\,G\over\hbar^{3/2}\sqrt{E_2} }
\end{equation}
\begin{align}
{d n_1(t) \over dt}
=
-{d n_2(t) \over dt}
=
\sqrt{n_2(t)}n_1(t)\left|\kappa \right|
\sin \left( \phi_2(t)-2\phi_1(t)-\phi_\kappa\right)
\label{dynPopulHighDens}
\end{align}
These expressions are similar to the equations describing the parametric interaction in second order nonlinear media~\cite{Armstrong1962} between the waves at frequencies $E_1=\omega$ and $E_2=2\omega$. The solutions are expressed using Jacobi elliptic integrals. Energy conservation implies that the exciton number times their energy $\sum_j E_j n_j=\sum_j E_j \left|\alpha_j\right|^2=\hbar \sum_j \left|A_j\right|^2$ is conserved. It results in
\begin{equation}
\left|A_1(t)\right|^2+\left|A_2(t)\right|^2=\left|A_0\right|^2
\end{equation}
The resonant case, when $E_2=2.E_1$, illustrates the main behavior of the two coupled oscillators. For an initial population of the $E_1$ oscillator, we get for example~\cite{Rosencher2002}
\begin{align}
	n_2(t)
	=
	\frac{n_1(0)}{2}\, \tanh^2 \left({t\over\tau_c}\right)
	& &
	n_1(t)
	=
	n_1(0)\,\left({1 - \tanh^2 \left({t\over\tau_c}\right)}\right)
\end{align}
that shows a full transfer (Fig.~\ref{Fig_OscClassiques}) of the populations from the $\rm E_{11}$ exciton described by the $E_1=\omega$ oscillator towards the $\rm E_{22}$ exciton that corresponds to the $E_2=2\omega$ oscillator within a time 
$
	\tau_c = {1/ \kappa \sqrt{n_0}|}
$

The coherence time of the excitons must be longer than $\tau_c$ to ensure that complete transfer of the populations, otherwise the oscillation of the two systems are randomly decoupled and the efficiency of the process decreases. $\tau_c$ goes to infinite when the initial exciton density $\left| A_0 \right|^2$ goes to zero. In the low density limit, one has thus to develop a quantum model for the nonlinear coupling of the $\rm E_{11}$ and $\rm E_{22}$ excitons.

As we will see, this will evidence a situation of strong coupling. Nevertheless the diagonalization of the Hamiltonian is delicate because the nonlinearity of the coupling term makes it depends on the exciton number. The transition probabilities are given by: 
\begin{eqnarray}
\left| \left\langle n'_1,n'_2\right| G^* . \,{\hat a}_1^\dag{\hat a}_1^\dag{\hat a}_2 \left| n_1,n_2 \right\rangle \right|^2 
& = &  
\left| G \right|^2 . \left( n_1+2\right) \left( n_1+1\right) n_2 . \delta_{n'_1 , n_1+2}\, \delta_{n'_2, n_2-1}
\label{asym1}\\
\left| \left\langle n'_1,n'_2\right| G . \,{\hat a}_1{\hat a}_1{\hat a}_2^\dag \left| n_1,n_2 \right\rangle \right|^2 
& = &  
\left| G \right|^2 .  n_1  \left( n_1-1\right) \left( n_2+1\right) . \delta_{n'_1 , n_1-2}\, \delta_{n'_2, n_2+1}
\label{asym2}
\end{eqnarray}
while in the case of a linear coupling between oscillators at the same frequency, we would get:
\begin{eqnarray}
\left| \left\langle n'_1,n'_2\right| g^*\,{\hat a}_1^\dag{\hat a}_2 \left| n_1,n_2 \right\rangle \right|^2 
& = &  
\left| g \right|^2 .  \left( n_1+1\right) n_2 . \delta_{n'_1 , n_1+1}\, \delta_{n'_2, n_2-1}
\\
\left| \left\langle n'_1,n'_2\right| g\,{\hat a}_1{\hat a}_2^\dag \left| n_1,n_2 \right\rangle \right|^2 
& = &  
\left| g \right|^2 .  n_1 \left( n_2+1\right) . \delta_{n'_1 , n_1-1}\, \delta_{n'_2, n_2+1}
\end{eqnarray}
Equations~\eqref{asym1} and~\eqref{asym2} reflect the asymmetry between the two oscillators for a nonlinear coupling: For example, when only the $E_2$ oscillator is populated ($n_2\neq 0$, $n_1=0$), the probability to transfer energy to the $E_1$ oscillator is proportional to $\left( n_2+1\right) $ while for the opposite case, when the $E_1$ oscillator is populated ($n_2= 0$, $n_1\neq 0$), the probability to transfer energy to the $E_2$ oscillator is proportional to the product $\left( n_1+2\right).\left( n_1+1\right)$.

\begin{figure}[h!]
	\begin{center}
		\includegraphics[scale=0.6]{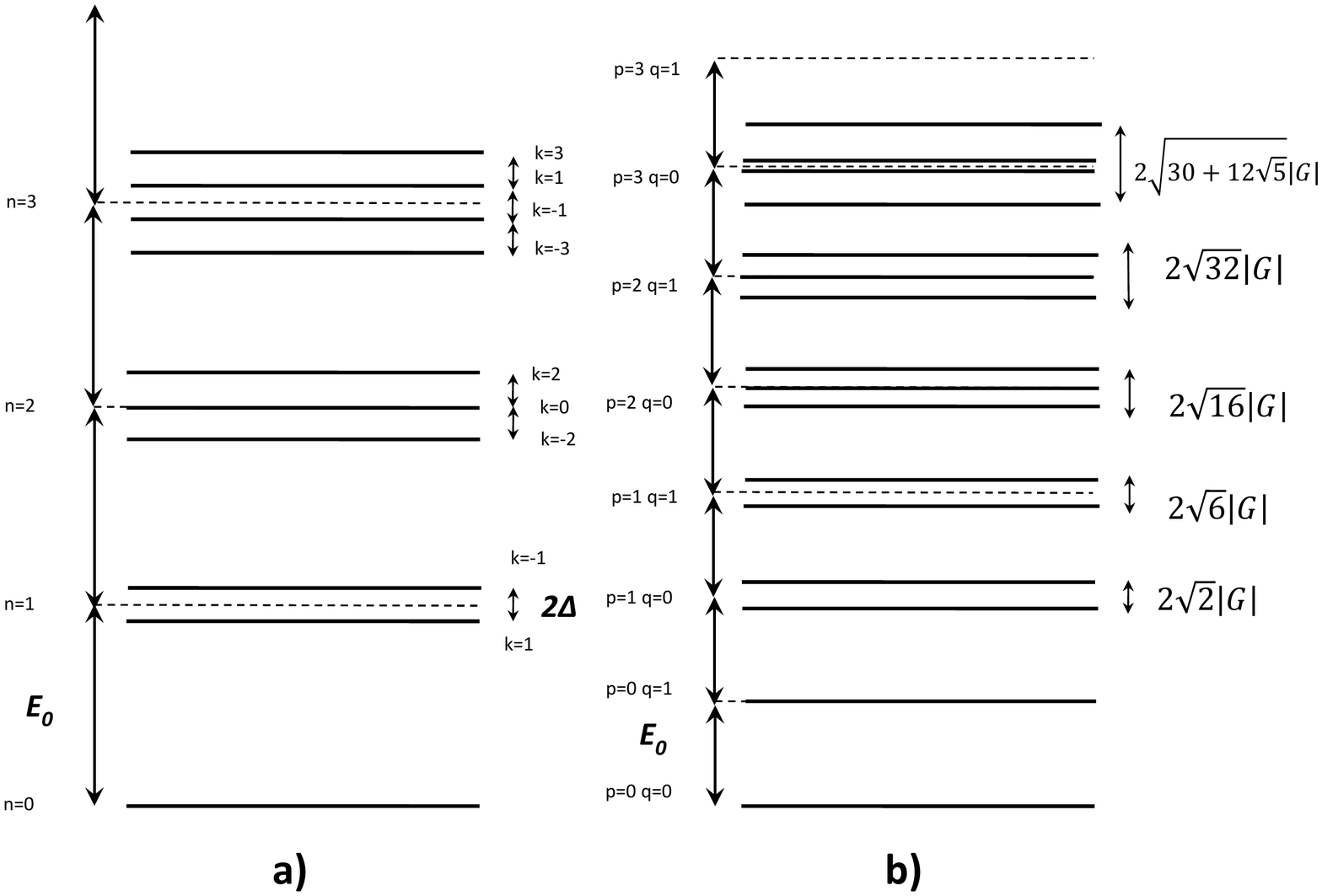}
		\caption {a. Linearly coupled oscillators.  A given total number $n= n_+ + n_- $ of quanta define a subset of the Fock space, with $n+1$ equidistant levels around $n.E_0$, that are separated by the energy $\Delta$ and labeled by the number $k = n_+ - n_- $. 
			b. Oscillators with a nonlinear coupling. For each subset around the energy $(2p+q).E_1$, there are $2p$ levels with a splitting that now depends on the numbers $p$ and $q$, contrarily to the linear coupling case.}
		\label{Fig_niveauxOscLin}
	\end{center}
\end{figure}

Before looking at the nonlinear coupling between harmonic oscillators, we can examine the linear coupling, for which the way to diagonalize the Hamilonian is well known, the polariton problem belonging to that case. The Hamiltonian can be written
\begin{eqnarray}
{\hat H}^{(L)}
& = & 
E_1\, \left({\hat a}_1^\dag {\hat a}_1 +{1\over 2}\right)
+E_2\, \left({\hat a}_2^\dag {\hat a}_2 +{1\over 2}\right)
+g^*\,{\hat a}_1^\dag{\hat a}_2
+g\,{\hat a}_1{\hat a}_2^\dag	
\\
& = & 
E_+\, {\hat B}_+^\dag {\hat B}_+ 
+E_-\, {\hat B}_-^\dag {\hat B}_-
\end{eqnarray}
with 
\begin{eqnarray}
{\hat B}_\pm^\dag
& = & 
u_\pm{\hat a}_1^\dag+v_\pm{\hat a}_2^\dag 
\end{eqnarray}
and
\begin{eqnarray}
E_\pm & = &  {E_1+E_2\over 2} \pm {1\over 2} \sqrt{\left(E_2 -E_1\right)^2+4 \left|g\right|^2 }
\end{eqnarray}
When $E_2\neq E_1$, the energy spectrum can divided into subsets, each corresponding to a given number $n$ of quanta (Fig.~\ref{Fig_niveauxOscLin}). With $n_\pm=\left\langle {\hat B}_\pm^\dag {\hat B}_\pm\right\rangle$, 
$n = n_+ + n_-$, and $k = n_+ - n_-$, we have
\begin{eqnarray}
E_{n,k} & = &  n_+ . E_+ + n_- . E_-
\\
& = & n.E_0 + k.\Delta  
\end{eqnarray}
where $E_0=\frac{E_1+E_2}{2}$ and $\Delta=\frac{\sqrt{\left(E_2 -E_1\right)^2+4 \left|g\right|^2}}{2}$. Each of these states is a linear combination of the uncoupled initial states with energies $n_1.E_1$ and $n_2.E_2$ and a given total number of quanta $n = n_1 + n_2$. An alternative method for diagonalizing the Hamiltonian would thus consist in doing it within each of those subspaces. 

We will use this method to handle the Hamiltonian of equation~\eqref{NLhamiltonian} $H_{NL}$ with a nonlinear coupling. For that case, each time a single quantum of the $E_2$ oscillator is created (resp. annihilated), two quanta are annihilated (resp. created) in the $E_1$ oscillator. Since only states with the same parity for $n_1$ are thus coupled, we can label the states using two indexes $p$ and $q$, where $q=0,1$ gives the parity. $n_1$ will go from its maximum $2.p+q$ down to $0$, when $n_2$ increases from $0$ to $p$. In this subset basis $\left\lbrace \ket{q, p} ,  \ket{q+2, p-1}, ... \ket{q+2j, p-j} ... \ket{q+2p, 0}\right\rbrace $, the Hamiltonian is a matrix with non-zero elements along the diagonal and on adjacent elements.
\begin{eqnarray}
	{\hat H}^{(NL)}_{j,j} & = & \left(q+2j\right).E_1 +\left(p-j\right).E_2
	\\
	{\hat H}^{(NL)}_{j,j+1} & = & \sqrt{\left(q+2j+1 \right) \left(q+2j+2 \right) \left(p-j \right) } .G 
	\\
	{\hat H}^{(NL)}_{j,j-1} & = & \sqrt{\left(q+2j-1 \right) \left(q+2j \right) \left(p-j+1 \right) }.G^*
\end{eqnarray}
As discussed in the following, we will be interested only in weak exciton numbers and we will examine only the first subsets. The lowest energy subset couples states $\ket{n_1=2,n_2=0}$ and $\ket{n_1=0,n_2=1}$, corresponding to $p=1$ and $q=0$ (Fig.~\ref{Fig_niveauxOscLin}). 
The diagonalization gives two energies
\begin{eqnarray}
E_\pm & = &  {2E_1+E_2\over 2} \pm {1\over 2} \sqrt{\left(E_2 -2E_1\right)^2+8\left|G\right|^2  }
\label{levelsplitting}
\end{eqnarray}
Two new states are symmetrically located with respect of the mean energy between two $\rm E_{11}$ excitons and one $\rm E_{22}$ exciton levels, with a splitting that depends on the coupling constant $G$. 
If we consider only the resonant case when $E_2=2.E_1$ and write $G=\left|G\right|e^{{\rm i}\phi}$. We get then $E_\pm=2E_1 \pm \sqrt{2}.\left| G\right|$ for the two states
\begin{equation}
\ket{\pm}=\pm\frac{e^{{\rm i}\phi}}{\sqrt{2}}\ket{n_1=0,n_2=1}
+\frac{1}{\sqrt{2}}\ket{n_1=2,n_2=0}
\label{mixedstate}
\end{equation}
As for the linear case, we get mixed state corresponding to a coherent coupling.
We notice here an equal weight for the two components. 
The next subset with $\ket{n_1=3,n_2=0}$ and $\ket{n_1=1,n_2=2}$, corresponding to $p=1$ and $q=1$ give similar results with $E_\pm=3E_1 \pm \sqrt{6}.\left| G\right|$ .

Things change for higher subsets, when more than two levels are coupled. For $p=2$ and $q=0$, three  states are involved: $\ket{n_1=4,n_2=0}$, $\ket{n_1=2,n_2=1}$, and $\ket{n_1=0,n_2=2}$. Energies are  
\begin{eqnarray}
E_\pm & = &  4E_1 \pm 4.\left| G\right|
\\
E_0 & = & 4E_1
\end{eqnarray}
with states
\begin{eqnarray}
\ket{\pm} & = & \frac{e^{{\rm i}2\phi}}{2\sqrt{2}}\ket{n_1=0,n_2=2}
\mp \frac{2e^{{\rm i}\phi}}{2\sqrt{2}}\ket{n_1=2,n_2=1}
+\frac{\sqrt{3}}{2\sqrt{2}}\ket{n_1=4,n_2=0}
\\
\ket{0} & = & -\frac{\sqrt{3}e^{{\rm i}2\phi}}{2}\ket{n_1=0,n_2=2}
+\frac{1}{2}\ket{n_1=4,n_2=0}
\end{eqnarray}
We see that the eigenstates are now asymmetric, with different weights for the two original oscillator states, as expected from the results for the classical oscillator described above. (This will give rise, for higher numbers of quanta, to the population transfer towards the $E_2$ oscillator). Nevertheless, as in the case of two linearly coupled oscillators, strongly coupled states are also formed. The main difference is that we do not obtain now eigenmodes that can be populated independently: the states and their energies depends on the populations.

This study of harmonic oscillators that have an energy ratio close to 2 and that are nonlinearly coupled, shows two important features: First, a strong coupling is observed, corresponding to coherent superposition of states of the two oscillators. Second, at high numbers of quanta, in the limit of classical oscillators, a full transfer of the energy towards the $E_2$ arises.

\section*{Exciton interaction}
To further develop our theory for excitons, we have now to look at their interactions. We write the exciton wave function as 
\begin{align}
\Phi^{(p,q)}_{K,n, J,M}(e, h)
&=
F_{n}( r )
.
{e^{i K . R} \over \sqrt{ L}} .u^{(p)}_c( r_e)
.
{u^{(q)}_{v}}^\ast( r_h)
\braket{JM}{j_e, m_e,j_h m_h}
\label{excitonwavefunction}
\end{align}
where $e$ et $h$ are shorthand notations for $\{ r_e, j_e, m_e\}$ and $\{ r_h , j_h, m_h\}$. $u^{(q)}_v$ and $u^{(p)}_c$ contains valence and conduction band periodic functions as well as envelope confined functions around a nanotube section. The respective minibands that are induced by the confinement in the transverse diameter of the nanotube are denoted by $q$ and $p$. $r_e$ and $r_h$ are the electron and hole position along the nanotube and $\{j_e, m_e\}$ and $\{j_h, m_h\}$ are their spin. $F_n(r)$ is the wave function of the electron-hole relative  motion. $K$ is the wavevector of the exciton; it appears in the exponential describing the wavefunction of its center of mass. We use the variables $r=r_e-r_h$ for the relative motion and $R$ for the center of mass. $JM$ denotes the total angular momentum of the exciton.

We simplify the notation by denoting  $i=\left\lbrace p,q ; K,n ; J,M \right\rbrace$. The creation operator of the exciton is given by 
\begin{align}
\hat{a}_i^{\dag}
&=
\iint de dh \, \Phi_i(e, h) \, \hat{\psi}^{\dag}\left(h\right)\hat{\psi}^{\dag}\left(e\right)
\end{align} 
where $\hat{\psi}^{\dag}\left(e\right)$ is the creation operator of an electron at point $r_e$ with spin $j_e m_e$ and $\hat{\psi}^{\dag}\left(e\right)$ the same for the hole.
Following~\cite{HaugSchmittRink1984}, we assume that excitons behave as quasi-bosons as described by the usual commutation rules  
\begin{eqnarray}
\left[
{\hat a}_i
,
{\hat a}_j^\dag 
\right]
& = &
\delta_{ij}+O(n_j)
\end{eqnarray}
where $n_j$ is the exciton density. The interacting electron and hole Hamiltonian is 
\begin{align}
\hat{H}
&=  
\int de \,
\hat{\psi}^{\dag}\left( e \right) 
T_e
\hat{\psi}\left( e \right) 
+\int dh \,
\hat{\psi}^{\dag}\left( h \right) 
T_h
\hat{\psi}\left( h \right) 
-\iint de dh \,
\hat{\psi}^{\dag}\left( h \right) 
\hat{\psi}^{\dag}\left( e \right) 
V_{eh}
\hat{\psi}\left( e \right) 
\hat{\psi}\left( h \right) 
\end{align}
and the wavefunction~\eqref{excitonwavefunction} is the Schrödinger equation solution in the single electron-hole case. For larger densities, interactions between excitons involve interactions between electrons $ee'$, holes  $hh'$, and electron and holes $e'h$ and $eh'$. With
\begin{align}
V
=
\frac{e^2}{4\pi }
\left(
-\frac{1}{4\pi\left| r_e-r'_h\right| }-\frac{1}{4\pi\left| r_h-r'_e\right| }+\frac{1}{4\pi\left| r_e-r'_e\right| }+\frac{1}{4\pi\left| r_h-r'_h\right| }
\right)
\end{align} 
we obtain
\begin{align}
I_D
=
\iiiint dedhde'dh'
{\Phi_4}^*(e, h)
{\Phi_3}^*(e', h')
\, V\, 
{\Phi_2(e', h')}
{\Phi_1(e, h)}
\end{align}
for the direct interaction term and 
\begin{align}
I_X
=
\iiiint dedhde'dh'
{\Phi_4}^*(e, h')
{\Phi_3}^*(e', h)
\, V\, 
{\Phi_2(e', h')}
{\Phi_1(e, h)}
\label{exchangeterm}
\end{align}
for the exchange interaction term~\cite{Ciuti1998}, where either the electrons or the holes are permuted. We will consider only the lowest exciton state $n$ for each mini-band $pq$. Taking into account that the electron and hole masses are the same in nanotubes, we obtain wave-functions that are symmetrical by permutation of $r_e$ and $r_h$. That gives rise to the mutual cancellation of direct terms like $I_D^{(eh')}$ and $I_D^{(ee')}$. Concerning now the exchange terms $I_X$, the integral implies an overlap between the wave functions in order to obtain a non-zero contribution. This means that the exchange interaction has very short range and is a contact interaction, over distances of the order of the exciton Bohr radius. 

This expression describes not only interactions within an exciton population sharing a specific state, but also processes like EEA where, on one side, two excitons belong to the $\rm E_{11}$ states and, on the other side, one obtains an exciton in the $\rm E_{22}$ stats and the annihilation of the last one, as described by the vacuum state $\ket{\phi}$ with $\braket{r_e r_h}{\phi}=\delta\left( r_e-r_h \right)$. The exchange term~\eqref{exchangeterm} becomes
\begin{align}
I_X
=
\iiiint dedhde'dh'
{\Phi^{(22)}}^*(e, h')
\, V\, 
{\Phi^{(11)}(e', h')}
{\Phi^{(11)}(e, h)}
\end{align}

We can now perform some approximations. In particular, we can neglect the dependency on the $K$-vector of both the exciton kinetic energy and their interactions: First, excitons that are optically created are concentrated around the bottom of the dispersion curve, close to $K=0$. Second, the kinetic energy of the exciton is very small if compared to the interaction energy, which amounts to consider excitons with infinite masses. Using the Fourier transform of the operators ${\hat a}_i$, that are the creation operators of an exciton ${\hat \Psi}^\dag\left( R \right)$ at point $R$, the exciton Hamiltonian can then be written as:
\begin{align}
\hat{H} 
= 
&    
\int dR \,  
\left(
E_{1} . {\hat \Psi}^{(11)\dag}\left( R \right) {{\hat \Psi}^{(11)}}\left( R \right) 
+ E_{2} . {\hat \Psi}^{(22)\dag}\left( R \right) {{\hat \Psi}^{(22)}}\left( R \right) 
\right.
\label{hamiltonienSpatial}
\\
& \left.
+ \frac{J_X}{2} . {\hat \Psi}^{(22)\dag} \left( R \right) {{\hat \Psi}^{(11)}}\left( R \right) {{\hat \Psi}^{(11)}}\left( R \right) 
+\frac{J^*_X }{2} .  {{\hat \Psi}^{(22)}}\left( R \right) {\hat \Psi}^{(11)\dag}\left( R \right){\hat \Psi}^{(11)\dag}\left( R \right)
\right)
\nonumber
\end{align}
where
\begin{align}
J_X=
\frac{1}{\sqrt{ L}}\iint dr\,dr' \,
F^*_{n_2}\left(  \frac{r + r'}{2} \right) 
.
\frac{-e^2}{4\pi \left|r\right| }
F_{n_1}( r' )
.
F_{n_1}( r )
\end{align}
with a normalization factor that involves the nanotube length $L$. We see that the structure  of that Hamiltonian is then similar to the one used in the previous part and we can take advantage of their conclusions: The Hamiltonian can be diagonalized for each given number of excitons and we obtain mixed $\rm E_{11}$ and $\rm E_{22}$ exciton states that are strongly coupled by the nonlinear interaction term. 

We can now discuss the strength of the coupling between $\rm E_{11}$ and $\rm E_{22}$. Coulomb interaction is known to be very efficient in carbon nanotubes. It gives an exciton binding energy that is the range of an eV. The interaction between excitons gives rise to the formation of biexcitons. Their binding energy was measured  to be $\rm 140\, meV$ in (6,5) SWCNT~\cite{Yuma2013, colombier_2012}. For trions, the binding energy is as large as $\rm 205\, meV$~\cite{Yuma2013}. The same term in the Hamiltonian gives rise to the strong coupling we evidence here. The $\rm E_{22}$ line is observed at 2.19~eV, i.e. 330~meV below the twice the energy of the $\rm E_{11}$ exciton (1.26~eV). This difference is usually attributed to the so-called trigonal wrapping, responsible for a curvature of the graphen energy dispersion (that is not anymore linear at high energies). The energy difference that we obtain here is nevertheless of the order of magnitude as the biexciton binding energy and must in a large part be due to the energy splitting (equation~\eqref{levelsplitting}) of the strongly coupled states.

\section*{Exciton spatial dimensions}
Equation~\eqref{hamiltonienSpatial} describes an Hamiltonian that is roughly separable into independent contributions for each position. It shows that two $\rm E_{11}$ excitons have to reach the same point to interact and to couple the $\rm E_{22}$ state. It is thus important to precise the different spatial features of the excitons~\cite{luer_size_2009}, that are often mistaken and deserve to be clarified. 

The excitons, in 1-D carbon nanotubes as well as in 2D-quantum wells or 3D-bulk crystals, show several specific spatial dimensions. From equation~\eqref{excitonwavefunction}, we see that two different scales characterize a Wannier exciton eigenstate: The electron-hole relative motion $F_{n}( r )$ has a spatial extension given by the Bohr radius $a_B$, while the center of mass wave function ${e^{i K . R} }/{ \sqrt{ L}}$ show that it is fully delocalized over the whole length $L$ of the nanotube. (Even  in the case of a Frenkel exciton in molecular system, where the excitations do not spread over several sites, with $a_B$ going to zero, the exciton wave function is not localized but, on the contrary, it covers the whole nanotube.) Nevertheless collisions processes induces dephasing and reduces quickly after their creation, the coherence length~\cite{Miyauchi_coherence_2009} of the center-of-mass wave-function down to a finite value $\ell_c$. In the limit of very frequent collisions, $\ell_c$ tends to zero and the exciton can be seen as a point-like particle. On the other side, particles with a finite size that encounter collisions show a diffusive motion that is characterized by a length $\ell_D$~\cite{cognet_stepwise_2007}. Both $\ell_c$ and $\ell_D$ have their time counter part with, respectively, the exciton coherence time $T_2$ and lifetime $T_1$.

\begin{figure}[h!]
	\begin{center}
		\includegraphics[scale=0.5]{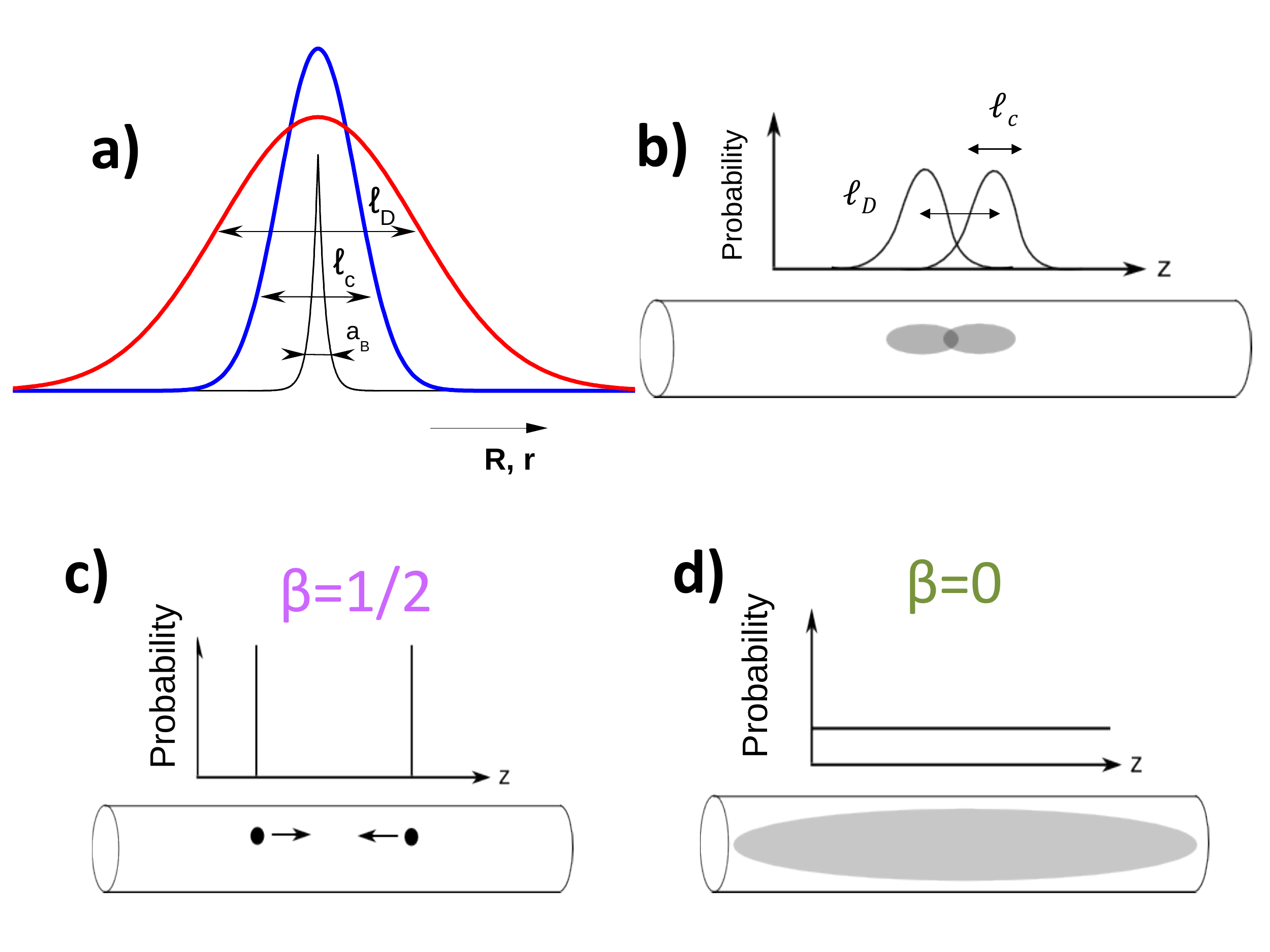}
		\caption {Finite size exciton
			a. and b. Schematic representation of exciton typical lengths: $\ell_D$ the diffusion length, $\ell_c$ the coherence length, and $a_B$ the Bohr radius. Excitons collide at very short distances, on a range given by $a_B$. They have a wave packet spatial extension given by $\ell_c$ and explore before recombining a spatial region with a size given by $\ell_D$. 
			c. and d. Limits of point like particles, diffusing in 1d extended systems and exciton in 0d quantum dots, whose wavefunction covers the whole crystal. $\beta=d/2$ is the corresponding parameter used in equation~\eqref{eqexcitondecay}.}
		\label{fig_spatial}
	\end{center}
\end{figure}

To summarize, excitons are characterized by three different lengths (Fig.~\ref{fig_spatial}a): The Bohr radius $a_B$, for the electron-hole relative motion (variable $r$); it gives as shown above the range of the exchange interaction; it is quite short and close to $\rm  2\, nm$ in (6,5) nanotubes. For the center-of-mass motion  (variable $R$), the diffusion length~$\ell_D$ gives first the spatial range that is explored by the exciton before it recombines; it was estimated to be close to $\rm 90\, nm$~\cite{cognet_stepwise_2007}. Second, the coherence length~$\ell_c$ can be estimated from the homogeneous spectral linewidth~\cite{Feldmann1987, Andreani1994}, knowing the exciton mass: One exciton is a wave-packet that is a superposition of states in the K-space. From the dispersion curve, the spread $\Delta K$ over K-vectors can be calculated and by Fourier transform, we get the real-space extension that is the coherence length $\ell_c$. From the linewidth of our absorption spectra $\gamma < \rm 20\, meV$, we obtain  $\ell_c > \rm 5\, nm $.

\section*{Exciton population and dynamics}
We have seen that the nonlinearly coupled oscillator behavior strongly depends on their population and we must evaluate now the number of excitons that are excited in SWCNT. Using the absorption cross-section value of $1\times10^{-17}{\rm cm}^2$ per carbon atom~\cite{berciaud_luminescence_2008} at low excitation, we can estimate maximum exciton density of $55 \rm \mu m^{-1}$ in the experiments we present below, \textit{i.e.} a separation of $18~\rm nm$ between two excitons~\cite{Yuma2013}. The mean distance between excitons is thus four times the size of their wave-packet, that is given by $\ell_c$.

That gives rise to relatively low exciton densities and the equations~\eqref{dynPopulHighDens} do not apply. Only the two first states with $n_1=1$ and $n_1=2$ are involved in the dynamics. As shown above, the short range of the interaction implies an overlap between two colliding excitons to get a transfer toward the mixed state described by~\eqref{mixedstate}. Such an overlap does not appear in equations~\eqref{dynNL} and we have to take it into account together with the spatial distribution of the excitons. When the mixed $2\rm E_{11}$-$\rm E_{22}$ state is formed, it can decay by two ways, in two processes based on the strong coupling between $2\rm E_{11}$ and $\rm E_{22}$ excitons that open a coherent ultrafast relaxation path between those two states: Either the $\rm E_{22}$ component decays for example by ionizing, as expected for creating the free carrier population that is evidence by the observation of trions in  photoexcited samples. Or the $2\rm E_{11}$ component dissociates into two
spatially separated excitons. Those processes dominate the first steps of the dynamics observed on the two excitonic resonances in pump-and-probe experiments~\cite{Langlois2015}. The rapid decay of the mixed state cross over the boundary between between reaction-and diffusion-limited kinetics~\cite{Allam2013} and gives rise to the creation of population of separated $\rm E_{11}$ excitons that shows an spatial antibunching.

On the opposite, two $\rm E_{11}$ which collides afterwards will be transfered to the mixed state that can decay efficiently by ionization at high energy. The dynamics of the population $n_1(t)$ of $\rm E_{11}$ excitons will thus be  described by an equation involving a bimolecular term like:
\begin{align}
\frac{d n_1(t)}{dt}=-C.n_1^2(t)
\end{align}
Nevertheless, this equation would valid for finite and small size systems such as quantum dots, where the excitons are coherent and overlap over the whole nanocrystal (Fig.~\ref{fig_spatial}c). For infinite systems, if considering colliding point-like excitons (Fig.~\ref{fig_spatial}b), one has now to take into account the diffusion of excitons~\cite{zhu_pump-probe_2007, murakami_nonlinear_2009, srivastava_diffusion-limited_2009} that have to reach the same point to interact. The equation turns then into 
\begin{align}
\frac{d n_1(t)}{dt}=-C.\frac{n_1^2(t)}{\sqrt{t}}
\end{align}
where the denominator takes into account the diffusive motion of the two particles before their collision. To describe real experiments, we must however consider an intermediate case, when the exciton wave-packet shows a finite coherent size (Fig.~\ref{fig_spatial}b). The excitons can interact at finite distances and the equation becomes
\begin{align}
\frac{d n_1(t)}{dt}=-C.\frac{n_1^2(t)}{t^\beta}
\label{eqexcitondecay}
\end{align}
with $\beta$ varying between 0 (exciton coherent over the whole crystal) and 1/2 (point-like exciton).

\begin{figure}[h!]
	\begin{center}
		\includegraphics[scale=0.5]{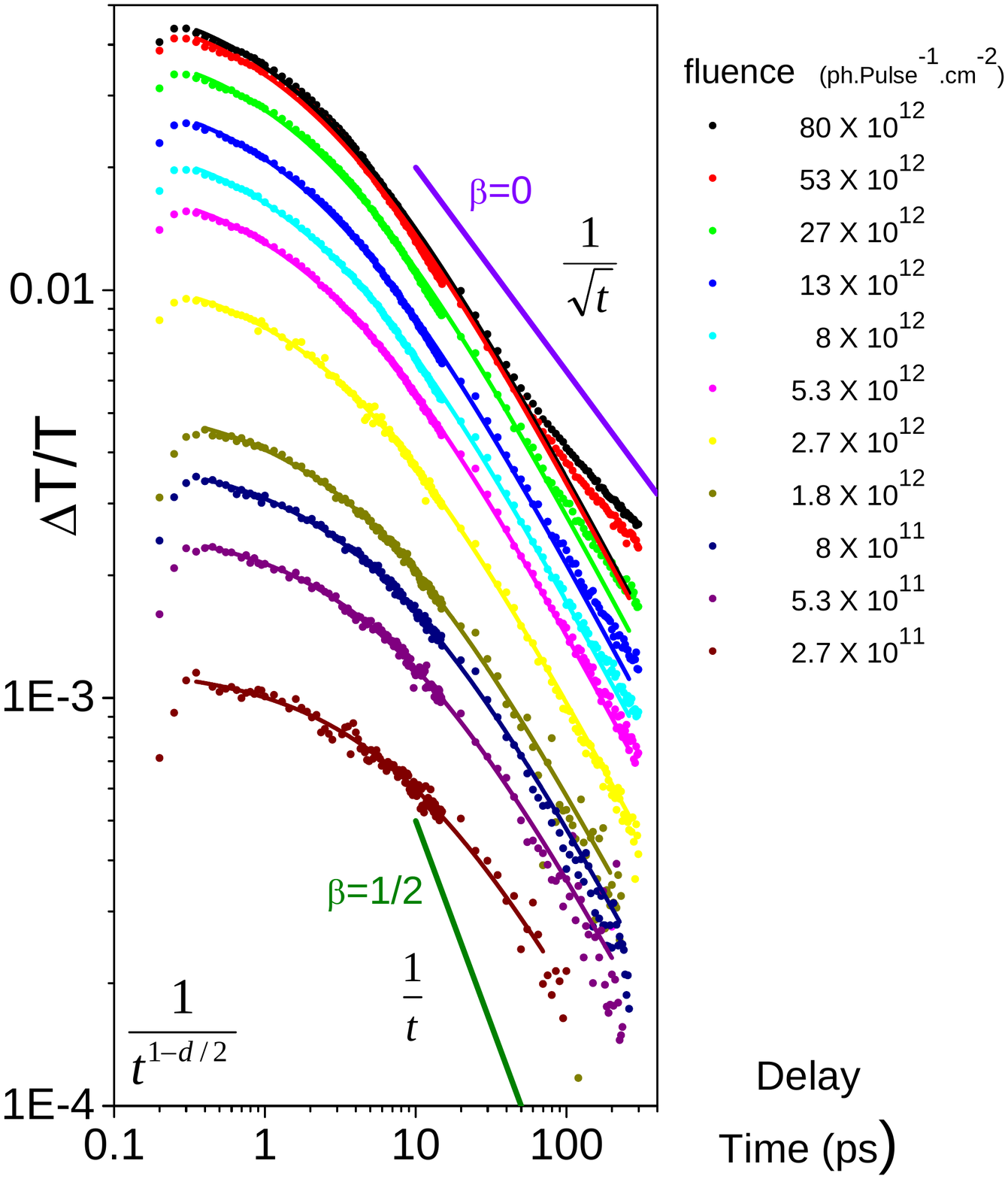}
		\caption {Intensity dependent transient dynamics of E11 exciton transmission in a suspension of chirality sorted (6, 5) SWCNT probed at 1.26~eV, following an excitation at 2.19~eV, resonant with the second-order (E22) optical resonance. Lines show fits of the exciton decay based on equation~\eqref{eqexcitondecay}.Two straight lines show the behaviors expected at long time delays for the values $\beta=d/2=1/2$ and $\beta=d/2=0$ of the parameter $\beta$ and of a system of dimension $d$.}
		\label{excitondecays}
	\end{center}
\end{figure}

To probe that behavior, we have measured the relaxation exciton dynamics in (6,5) nanotubes by performing pump-and-probe experiments. A detailed description of the experimental setup we used can be found elsewhere~\cite{Yuma2013}. The SWCNTs in solution are excited resonantly on the $\rm E_{22}$ transition. The absorption change at the $\rm E_{11}$ transition is measured as a function of the delay between the pump and the probe pulses. This gives a measurement of the $\rm E_{11}$ excitons decay~(Fig.~\ref{excitondecays}). Measurements performed at different excitation intensities allow to follow, for different initial exciton densities, the evolution of their dynamics. We see first that the initial changes are instantaneous, on the time scale of the pump pulse (100 fs), as expected for the excitation of coupled $\rm 2E_{11}-E_{22}$ states. We see next on this figure that curves are well fitted by solutions of the equation~\eqref{eqexcitondecay}
\begin{align}
n(t)=\frac{n_0}{1+C.n_0.t^{1-\beta}}
\end{align}
where $n_0$ is the initial exciton density and $C$ is a constant number.
We emphasize that this fitting procedure works, first, for each of the curves that extend on 3 orders of magnitude for the time delay (from $\rm 100\, fs$ up to $\rm 300\, ps$) and more than one order of magnitude for the signal intensity and, second, it works also for the whole set of curves that covers an excitation intensity range of more than 2 orders of magnitude. 

We find that the parameter $\beta$ is equal to $0.3$. It represent the random motion of interacting particles that have to reach the same point to collide. Equation~\eqref{eqexcitondecay} depicts usually such processes, with $\beta=d/2$ where $d$ is the dimension of the space. We find here a fractional dimension $d=0.6$, in between the two limiting cases of a quantum wire ($d=1$) and a quantum dot ($d=0$). This is due to the finite size of the colliding particle: it shows that they have a finite coherence length $\ell_c$. This is related to the spatial extension of the exciton center-of-mass wave function. It characterizes the overlap between the colliding excitons, even if their interaction, which is due to electron-hole exchange, occurs on a very short spatial scale, given by their Bohr radius. This coherence length gives also the spatial range where exciton complexes like biexcitons are formed, with a correlation between bound charges, while for larger distances only an incoherent interaction between those charges occurs.  

\section*{Discussion}
We can now give a comprehensive picture of excitons in SWCNTs and of their dynamics.

Because of the resonance between the energy of the second sub-band and twice the energy of the first sub-band, the nonlinear interaction between the $\rm E_{11}$ and $\rm E_{22}$ excitons gives rise to their strong coupling and to the formation of mixed states. It means that, if the lowest resonance is well the $\rm E_{11}$ exciton line, the second resonance is not the pure $\rm E_{22}$ exciton absorption but the signature of the mixing between the two states, which is pushed away towards lower photon energies by the splitting due to the interaction (Fig~\ref{Fig_strongcoupling}b).

The exciton population dynamics is as well clarified. While in the usual picture involving EEA, the final state belongs to the $\rm E_{11}$ manifold, we show that the exchange Coulomb interaction between $\rm E_{11}$ excitons couple them with the $\rm E_{22}$ state. The existence of the resulting strongly coupled mixed state explains some experimental features that have been reported: The ultrafast relaxation that is observed when the excitation is tuned into the $\rm E_{22}$ resonance and the fact  that a signal is instantaneously observed on the $\rm E_{11}$ resonance. 

Mixing if the two states gives also rise to the decay of the exciton population: as soon as the two $\rm E_{11}$ excitons forming the mixed state move away from each other in the nanotube, the coupled state is broken and the two excitons can decay radiatively. Nevertheless, the inverse process is more very efficient to decrease the exciton population. When two $\rm E_{11}$ excitons collide, they form a mixed state with the $\rm E_{22}$ level that is at high energy. The ionization of the exciton becomes more probable and it can lead to the expel of one carrier (electron or hole) into the environment of the nanotube and leave behind a single carrier: this leads to the observation of trions under high intensity excitations~\cite{Yuma2013}. Together with the presence of extrinsic defects, the strength of the exciton-exciton interaction, as an intrinsic effect, could thus explain the low radiative efficiency of the carbon nanotubes: When high densities of excitons are created, they collide rapidly and the transfer towards the higher states favorites further their dissociation and their non-radiative annihilation. This relaxation channel for the $E_{11}$ is coherent since it does not involve a transfer of population between the states, but their coherent superposition in a mixed state.

This scenario also explains why biexcitons are not observed in photoluminescence experiments: the very efficient nonlinear coupling between the two-$\rm E_{11}$ exciton state and the $\rm E_{22}$ exciton makes the biexciton formation by collision between two  $\rm E_{11}$ excitons very unlikely. 

We can also give some estimations of the typical lengths that characterize excitons in SWCNT. Their interaction is due to the exchange term and is thus of very short range, on the order of the Bohr radius (2nm). They form wave packets with finite spatial coherence length, in between the limit of point like particles, moving in extended systems, and excitons in quantum dots, whose wave-function covers the whole crystal. They can thus be seen as moving in a space with a fractional dimension $d=2.\beta=0.6$, value between the limits $d=1$ of a nanowire  and $d=0$ for a quantum dot.

\section*{Conclusion}

We have discussed here the dynamics of excitons in SWCNTs and their coupling. We have proposed the two-$\rm E_{11}$ state and the $\rm E_{22}$ state are strongly coupled, giving a very efficient coherent channel of energy exchange between high and low energy levels. These conclusions go beyond the case of the excitation dynamics in SWNTs: Strong nonlinear coupling should be present in many systems as soon as two conditions are fulfilled. If we consider boson-like excitations (that can be basically described as quanta of harmonic oscillators),  one needs: first a coincidence between the energy $\omega_b$ of one of the oscillators and the energy of an harmonic for the second one $n.\omega_a=\omega_b$, $n$ being an integer; second, an interaction should give rise to a nonlinear coupling. We can notice that the same results should be valid for anharmonic oscillators that are linearly coupled. This could have important consequences for the efficiency of many photonic systems. For example, in materials that are developed for photovoltaic solar cells, if excited state absorption is expected, the coupling with high energy states of two colliding exciton states should be as well efficient to limit the number of photogenerated carriers.

\bibliographystyle{unsrt}
\bibliography{SCNT}

\end{document}